\documentclass[conference]{IEEEtran}
\usepackage{graphicx}
\usepackage{cite}
\graphicspath{ {images/} }
\usepackage{amsmath} 
\usepackage{color,colortbl}
\usepackage{algorithm}
\usepackage[noend]{algpseudocode}
\usepackage{todonotes}
\usepackage{caption}
\usepackage{graphicx, subfig}
\usepackage{dblfloatfix}
\usepackage{comment}
\usepackage{authblk}
\usepackage[algo2e]{algorithm2e} 

\begin{document}
\title{A Software-Defined Approach for QoS Control in High-Performance Computing Storage Systems}

\author[1]{Neda Tavakoli}
\author[1]{Dong Dai}
\author[2]{John Jenkins}
\author[2]{Philip Carns}
\author[2]{Robert Ross}
\author[1]{Yong Chen}
\affil[1]{Computer Science Department, Texas Tech University, USA \{neda.tavakoli, dong.dai, 
yong.chen\}@ttu.edu}
\affil[2]{Mathematics and Computer Science Division, Argonne National 
Laboratory, USA, \{jenkins, carns,rross\}@mcs.anl.gov}

\maketitle
\IEEEpeerreviewmaketitle
\section{EXTENDED ABSTRACT}

High-performance computing (HPC) storage systems become increasingly critical to 
scientific applications given the data-driven discovery paradigm shift. 
As a storage solution for large-scale HPC systems, dozens of applications share the same storage system, and will compete and can interfere with each other. Application interference can dramatically degrade the overall storage system performance. Therefore, developing a flexible and effective storage solution to assure a certain level of resources per application, i.e. the Quality-of-Service (QoS) support, is critical. 
One of the common solution to achieve QoS assurance for storage systems is using provisioning technique~\cite{3}. Provisioning refers to the ability of providing certain amount of resources for applications and expected workloads. However, provisioning has limitations such as requiring the detailed knowledge of the expected workloads. In addition, the storage workloads are transient hence expensive to be satisfied. Due to these limitations, providing QoS storage systems through provisioning is challenging.

In this research, a software-defined approach~\cite{0} is proposed as a flexible solution to achieve QoS guarantee for storage systems. The driving force of using a software-defined approach instead of the traditional approaches, is that it has the ability to enable a more flexible, scalable, and efficient platform. For example, if any changes occurred in the system, it does not necessarily need to re-configure thousands of devices; instead, with re-configuring a logically centralized component, other devices will be automatically notified.

Therefore, in this research we focus on studying software-defined methods to meet the QoS requirements. 
Specifically, {\em a new software-defined approach is proposed to guarantee ``soft'' QoS for applications}. 
The framework allows the desired bandwidth to be specified by each application. Based on these specifications, applications are able to obtain their specified bandwidth if possible; or still get appropriate shares of the bandwidths even the storage systems are overloaded (i.e., a soft QoS guarantee).

\subsection{Software-Defined Approach for Storage System QoS}

\begin{figure}[htbp]
\begin{center}
{\includegraphics[width=3.2in]{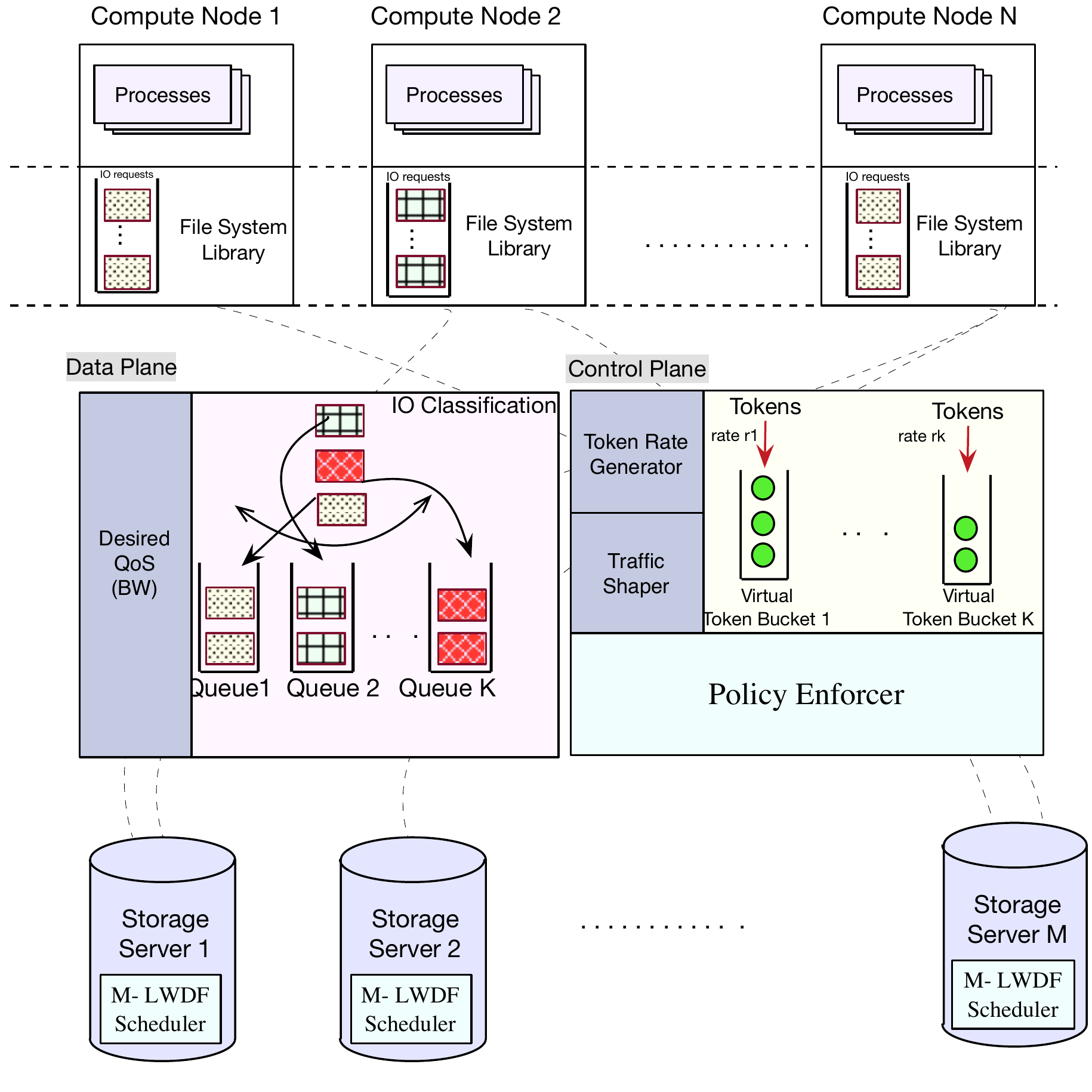}}
\caption{A software-defined approach for soft QoS control in HPC storage systems.}
\label{fig1}
\end{center}
\end{figure}

Fig.~\ref{fig1} shows the proposed architecture using software-defined technique. It consists of several components. An application runs on compute nodes in parallel.
On each compute node, there is a queue keeping IO requests from applications.
We add two key software-defined components into the HPC systems to enable a flexible QoS guarantee. They are \emph{data plane} and \emph{control plane} as the figure shows. Data plane runs on each storage server for IO classification and bandwidth shaping for each application. It contains multiple queues, each of which buffers requests from a given application. IO classification is done based on the IO header. IO requests with the same IO header belong to the same application and will be put to the same queue (in the data plane). As each application has a desired bandwidth requirement, we keep these specifications in \emph{Desired QoS} component in the data plane.

The control plane consists of several components. 
First, the \textit{Token Rate Generator} communicates with the \emph{Desired QoS} component in data plane to sync the requested bandwidth specification of each application. Based on these information, it generates a token rate per application and equally distributes those tokens into the queue of each corresponding application in the data plane. Token is a conceptual data structure representing the permission of performing IO requests. Only if a queue holding enough tokens, it will be served by storage system. Second, the \textit{Virtual Token Buckets} component learns the token rate from token rate generator to operate. Note that, the number of generated tokens per application is different. The \textit{Traffic Shaper} component communicates with the virtual token buckets to get information to shape the traffic. Last, the \textit{Policy Enforcer} is used to deliver policies to meet the QoS requirements. For example, a  policy can be \texttt{<app-1, rate=100 MB/s>}. Using this component for meeting the QoS requirements provides the flexibility to configure the system. It will apply the configuration policies to hundreds of storage servers. 

\subsection{Challenges}
Even though the proposed architecture provides flexible mechanism and traffic shaping capability to meet the QoS requirements, it still has two major challenges, and this study attempts to address them.

\subsubsection{\textbf{Unbalanced IO requests}}
\begin{figure}[htbp]
\begin{center}
{\includegraphics[width=2.3in]{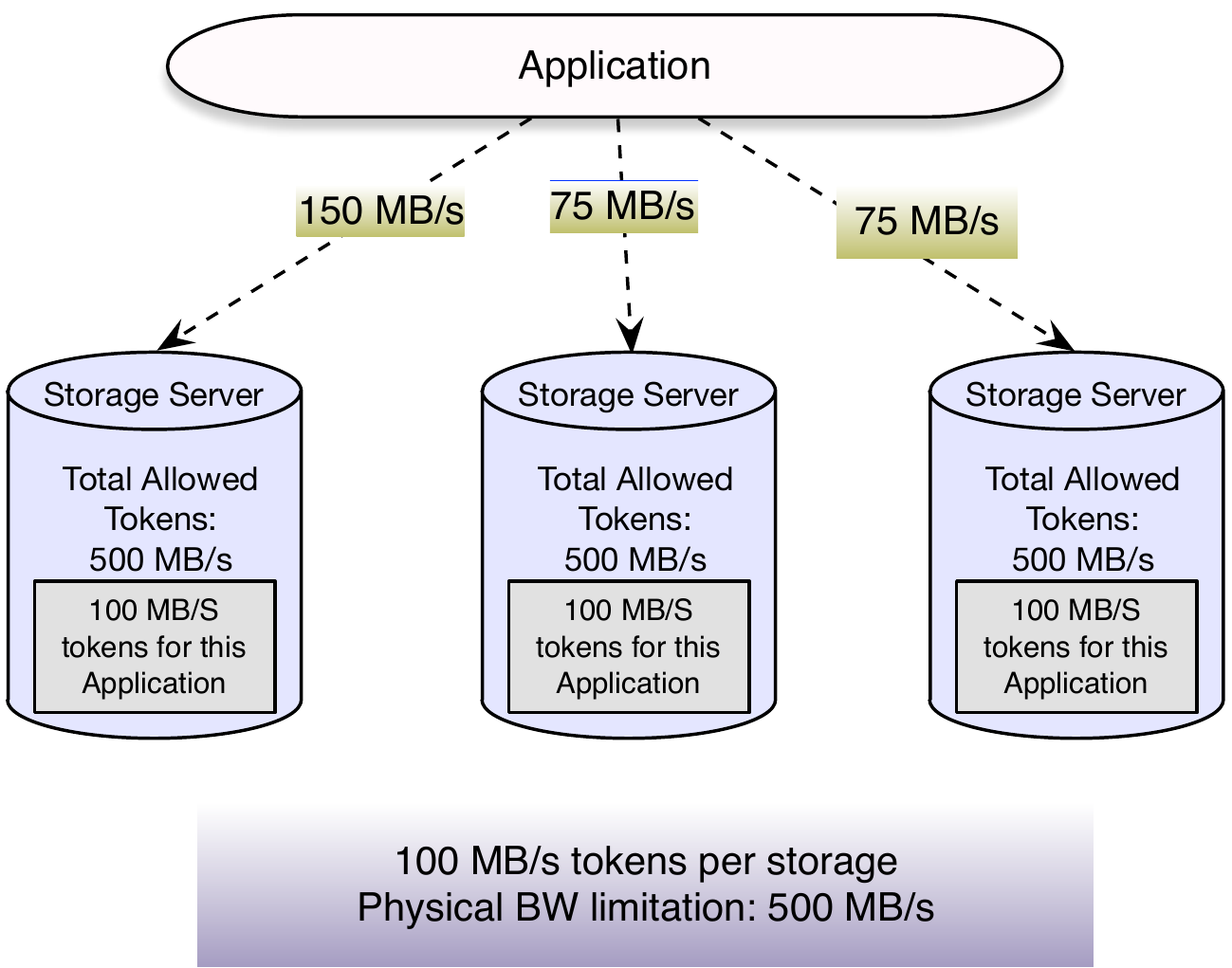}}
\caption{Unbalanced IO requests.}
\label{fig2}
\end{center}
\end{figure}

HPC applications often perform IOs in an unbalanced way, which can cause a challenge for the fair sharing of tokens among storage servers. For example, Fig.~\ref{fig2} shows how this could happen. For simplicity, we only take one application as example. It issues unbalanced IO requests from different compute nodes hitting different storage servers. Suppose this application requests 300 MB/s bandwidth QoS, the total physical bandwidth limit for each storage is 500 MB/s, and the control plane assigns 300 MB/s tokens for this application. These tokens are evenly distributed to storage servers such that each of them has 100 MB/S tokens.
As this application only contains 100 MB/s tokens in each storage, server 1 could not serve requests at 150 MB/s, even though other servers actually have unused tokens for this application. In this example, only a total 250 MB/s bandwidth is provided even there are unused tokens for this application. To address such an issue, we propose a borrowing model described in the next section.

\vspace{-1em}

\subsubsection{\textbf{Physical limitation}}
Each storage server has a physical bandwidth limitation. This limitation can be reached due to unbalanced IO requests from a single application or due to the concurrent data accesses from multiple applications. If there is a need exceeding the limitation, even though there are tokens available, QoS requirement can not be satisfied. When this happens, the desired QoS of those applications should degraded. The degradation should be fair among multiple applications. 

\subsection{Proposed Solution} 
In order to overcome these limitations, in this research, we propose a borrowing model in conjunction with extended {M-LWDF} algorithm~\cite{1,2} 
to meet the soft QoS guarantee. 
The borrowing model is a mechanism that allows a queue $Q_i$ for application $i$ to borrow tokens from queues of the same application in other storage servers. The model covers whether the borrow can happen, when and how many tokens should be borrowed. 
We also extend the original M-LWDF algorithm to guarantee the fairness during degradation. The original algorithm is widely used to provide certain minimum bandwidth for each application. It uses heuristic metrics to choose one application to serve each time. But it can only be used with a single shared resource (i.e., a storage server in our case). We extend this algorithm with a borrowing model, which introduces a dynamic number of tokens for each application, i.e., if $Q_i$ on server $S_i$ borrows extra tokens from other servers, it will gain higher priority to be served through M-LWDF algorithm. Once the hardware limitation is reached, it gets higher portion of bandwidth.

In addition, we design a set of policies regarding the borrowing model as follow:
\begin{itemize}
\item Prohibit an application to borrow tokens: \texttt{<app-i, borrow=FALSE>}.
\item Allow an application to borrow tokens: \texttt{<app-i, borrow=TRUE>}. 
\item Allow an application to borrow tokens if only $thres$ percent of its required bandwidth is satisfied: \texttt{<app-i, borrow=TRUE, thres=0.8>}.
\end{itemize}

Policy enforcer in the control plane will distribute the policy that users specify to control the QoS of an application.



\subsection{Summary}
In this research, we introduce a new software-defined approach to implement QoS control in HPC storage systems. We introduce the architecture, identify critical challenges, and propose a new borrowing model to address challenges. We expect to present early evaluation results at the conference.


\end{document}